# Dehydration, Dissolution and Melting of Cyclodextrin Crystals

Erika Specogna[(1)], King Wo Li[(1)], Madeleine Djabourov*[(1)], Florent Carn[(2)] and Kawthar Bouchemal[(3)]

(1) Laboratoire de Physique Thermique, ESPCI ParisTech, PSL Research University, 10 Rue Vauquelin 75231 Paris Cedex 5 France

(2) Laboratoire Matière et Systèmes Complexes, UMR 7057 Université Denis Diderot-Paris 7, Case Courrier 7056 10, rue Alice Domon et Léonie Duquet 75205 Paris Cedex 13, France

(3) Institut Galien Paris Sud, UMR CNRS 8612, Faculté de Pharmacie, University Paris-Sud 5 rue Jean-Baptiste Clément 92296 Chatenay-Malabry Cedex France

## Abstract

Cyclodextrins are a family of oligosaccharides with a toroid shape which exhibit a unique ability of entrapping guest molecules in their internal cavity. Water is the primary guest molecule and is omnipresent in the crystalline phases stabilizing the overall architecture. Despite the presence of water molecules inside the cavity, cyclodextrins provide a hydrophobic environment where poorly soluble molecules can easily fit. In this investigation we put in evidence different types of water in the hydrated α, β and γ cyclodextrin crystals. Thermogravimetric measurements identify various binding sites of water and highlight the difference between the crystals equilibrated under various humid atmospheres. We establish by microcalorimetry the limit of solubility versus temperature and measure for the first time the melting temperatures of the hydrated crystals. Dissolution and melting enthalpies are derived and the solubility curves are compared to existing literature. The specific features of each cyclodextrin are underlined.

# 1 Introduction

Cyclodextrins (CDs) are cyclic oligomers of D-glucose produced by action of certain microbial enzymes on starch. α, β and γ-CDs are composed of respectively 6, 7 and 8 D(+)glucopyranose units connected by α(1,4)-linkages. The molecules have the shape of a truncated cone. The largest diameter of the internal cavity varies according to the number of glucose units, being respectively 5.7 Å, 7.8 Å and 9.5 Å, for α, β and γ-CD, while the height of the truncated cone is *circa* 7.8 Å, for the three cyclodextrins. The CD crystals have been investigated independently in great detail by X ray diffraction [1–6] techniques. CD molecules are well known to form inclusion complexes in water with a wide variety of guest molecules, both organic and inorganic (see for instance references in [7]). While hydroxyl groups occupy both rims of the cone, the inside of the cavity is hydrophobic in character [8]. The cavities provide a hydrophobic matrix in aqueous solution, where the guest molecule should fit in even if this is only partially [9]. The inclusion mechanism involves weak interactions between included molecules and hydroxyl groups of CD. CDs are normally used to increase the apparent solubility [10] of poorly water soluble drugs. Moreover, polyrotaxanes with α-CD were prepared via a threading mechanism by inclusion of (poly-ethylene)glycol chains of various molecular weights or using other polymers [11,12]. Inclusion complexes may assemble in well-organized crystalline structures when synthetized under suitable conditions, as shown by X ray and neutron diffraction studies [7,13–20]. The packing mode depends on the guest molecule and the type of CD. The molecular packing in crystals is governed by the arrangement of hydrated CD molecules. The soluble complexes have been mostly analyzed with high resolution NMR techniques. It appears from the literature, that the degree of hydration of the "pure" crystals (water being the guest molecule of CDs) and of inclusion complexes (with other guest molecules) is very sensitive to the synthesis and drying protocols. The water content of the crystals can be tailored by quick precipitation of the complex or by sorption/desorption mechanism and the crystal structure will be modified accordingly. It is therefore important to understand the influence of hydration state on the crystalline structure and also on other properties - such as the thermal stability of crystals, dissolution properties in water or other solvents, maximum solubility - starting with water as a guest molecule and further on, with inclusion complexes. Surprisingly, such issues have been less raised in the literature for hydrated CD crystals. Jozwiakowski and Connors, [21] noticed that the solubility limit of CDs are rather low for saccharides. They performed equilibrium studies for dissolution with well controlled thermal conditions and established a solubility Van't Hoff diagram. More recently, Bastos et al. [22] studied the difference between enthalpies of dissolution of dry and hydrated solid samples of α-CD and established that the dissolution of hydrated α-CD is endothermic, while dissolution of dry α-CD is exothermic (-60 kJ/mole). In the latter case, the complex formation between water and "empty" CD is strongly exothermic. Therefore, the endothermic peak obtained for hydrated CDs dissolved in water is the result of the dissolution process, whereas the exothermic value for the dry compound reflects the sum of the two processes, complex formation with water molecules and dissolution. The presence of water molecules in the crystal is therefore an important parameter which discriminates between exothermic and endothermic processes of solubilization. Other experimental work on thermal properties of CDs performed by thermogravimetric analysis (TGA) [23] showed a stepwise dehydration with distinctly different energy levels of water. Sabadini et al. [24] also studied different steps of dehydration of α-CD hydrates, using both $H_2O$ and $D_2O$ as solvents. Rusa et al [20] identified the decomposition temperature of α-CD hydrates close to 256°C and the melting temperature is usually considered to be above 278°C.

In this paper we investigate the dehydration and dissolution of α, β and γ-CD solid hydrates by TGA and microcalorimetry (µDSC) under a large range of CD concentrations and temperatures. Prior to TGA measurements CD powders were equilibrated under controlled humidity atmospheres for 5 days. For dissolution studies the powders were used as received. We establish the solubility diagrams of CDs versus temperature, the dissolution enthalpies for α-CD and β-CD and for the first time to our knowledge we measure the melting temperatures of the CD crystalline hydrates and the corresponding melting enthalpies.

## 2  Materials and methods

The α-CD was produced by Wacker Chemie AG (lot BCBH0018V) purity ≥ 98 % and was supplied by Sigma-Aldrich. β-CD was purchased from Fluka Chemika (assay >99%) and γ-CD (Cavamax W8 Pharma) was produced by International Specialty Products (ISP). The powders were used without any further purification. Deionized water from Milli Q system with a resistivity of 18 MΩcm was used for CD powders solubilization. The solutions were prepared by stirring the dispersions with a magnetic bar at 25°C. We prepared aqueous solutions with concentrations measured by weight. We established that the optical angle increased linearly for α-CD solutions with concentration using a Perkin Elmer 341 polarimeter with Hg lamp at different wavelengths of 365, 436, 546 and 578 nm. The smallest wavelength gives the largest optical rotation angle. This method was used to establish the limit of solubility of α-CD after centrifugation of concentrated dispersions at 25°C. The centrifugation was performed with Eppendorf 5702RH centrifuge at 3000 rpm in falcon tubes during 30 min at 25°C. The data (in Supporting Information) shows the limit of solubility for α-CD powder at 25°C is *circa* 13.5 wt. %.

Prior to TGA measurements, CD powders were equilibrated in three desiccators at *RH*= 90%, 44 % and 3% in small Al dishes, during 5 days, at room temperature. The *RH* in desiccators was controlled by supersaturated solutions of respectively $BaCl_2$ and $K_2CO_3$ and by dry silica gel particles.

TGA was performed with Q 5000 from TA Instruments. Samples were placed in an open platinum pan 100µL that was hung in the furnace. The initial weight of the powder was around 30 mg and Nitrogen was used as the purge gas at a fixed flow of 25 mL/min. The weight of material was recorded during heating from 25 to 120 °C at a heating rate of +1°C/min. The heating ramp started immediately after the sample was placed in the Pt pan.

Water content (wt. %) of the powders determined by TGA for the three CDs after equilibration at different *RH*s and their initial values (as received powders) are given in Table 1.

Table 1

Water content (wt. %) of the CD powders at different *RHs* and in their initial states

| *RH*  | initial | 90%    | 44%    | 3%    |
|-------|---------|--------|--------|-------|
| α-CD  | 9.68%   | 10.15% | 10.11% | 5.26% |
| β-CD  | 13.59%  | 15.21% | 14.26% | 2.95% |
| γ-CD  | 9.03%   | 18.18% | 9.64%  | 4.55% |

MicroDSC (µDSC) experiments were performed with µDSC3evo from Setaram (Caluire, France) in batch Hastaloy cells by weighing suspensions with different concentrations from to $c$=10 to $c$=90 wt. %. The temperature ramp started at 25°C after thermal equilibration and the heating was performed with a constant rate of +0.1°C/min. The heating rates for TGA and µDSC experiments were as slow as possible to be compatible with the instrumental imitations.

X ray diffraction (XRD) patterns were recorded using a Panalytical Empyrean set-up equipped with a multi-channel detector (PIXcel 3D) using Cu-Kα radiation in the 3–50° range, with a 0.025° step size and 30s per step.

Scanning electron microscopy (SEM) was performed using a JSM-6510 microscope from JEOL equipped with tungsten filament. All the crystal' morphology observations were done with an acceleration voltage of 10 Kev. In supporting material we show the SEM images of the crystals of CD as received and after crystallization in water at room temperature for supersaturated solutions (concentrations >20 wt.%) of α-CD and β-CD. Observations of 'as–received' CD reveal that all the samples have crystalline morphologies with variation in the size distribution. α-CD and γ--CD are composed of small crystals with typical size around 10 µm while the raw β-CD powder is composed of large crystals with typical size around 100 µm. As anticipated, the recrystallization enables to obtain large crystals which can reach the millimetre size with smooth surfaces and cutting edges.

## 3  Results and discussion

### 3.1  Crystal structure

We investigated the structure of CD hydrates by X-ray diffraction. The patterns of "as received" crystals and those of α -CD recrystallized in a supersaturated aqueous solution, at room temperature, are shown in Figure 1.

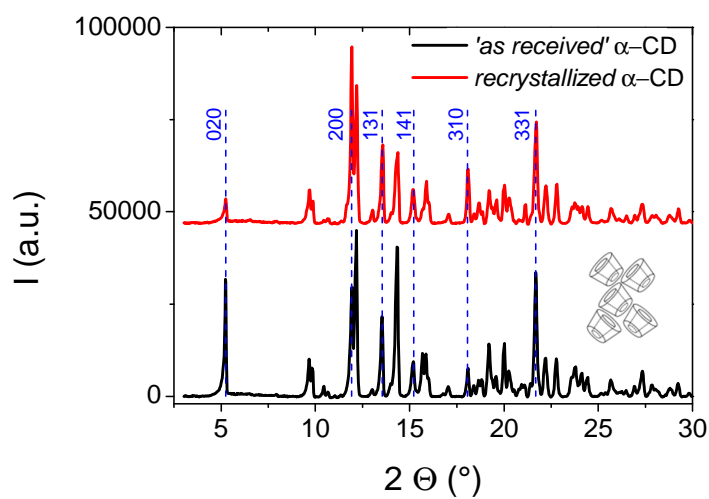

Figure 1 X-Ray diffraction patterns of "as received" α-CD (bottom, black curve) and recrystallized α-CD (top, red curve). The blue lines (dotted lines) indicate the position of the main diffraction peaks corresponding to the hydrated "cage" structure as reported by [2,24,25]

The molecular packing of α-CD in crystals is mainly known with two types arrangements: of the so-called "cage" ( or "herringbone") structure and the columnar structure (see for instance [8]). The characteristic diffraction peaks of our hydrated crystals correspond to the "cage" structure signature in good agreement with previous structural indexations performed by [2,24,25]

The degree of hydration is a very important factor and affects the crystal structure. Our α-CD samples ("as received") have a residual water content of *circa* 10 wt. % as discussed in the TGA section. Hunt et al. [25] showed WAXS diffraction measurements that the diffraction pattern of vacuum dried α-CD crystals changes during sorption experiments. During storage in humid atmospheres, upon vapor sorption, the "columnar structure" (initial structure) progressively converts into the "cage" structure. The sample investigated by Hunt et al was synthetized by precipitation of a clear solution with chloroform at room temperature, and the precipitate was immediately vacuum-filtered and dried overnight under a vacuum draft. Our results demonstrate that the crystals slowly formed in the supersaturated solutions at room temperature belong to the "cage" structure and the "as received" α-CD crystals with a water content of *circa* 10 wt. % are also in the same structure (Figure 1).

## 3.2 TGA

We investigated the dehydration of CD powders after storage in desiccators with controlled humidity. The three CDs were investigated. These experiments determine accurately the total initial water content of the powders after storage in controlled atmospheres and the dehydration steps related to different sites of the water molecules in the hydrated crystals.

### 3.2.1 α-CD

Figure 2a shows that the total water content is quasi identical for "as received" crystals and after equilibration with $RH$=44% and 90%. A significant amount of water however desorbed in the dry atmosphere ($RH$=3%).

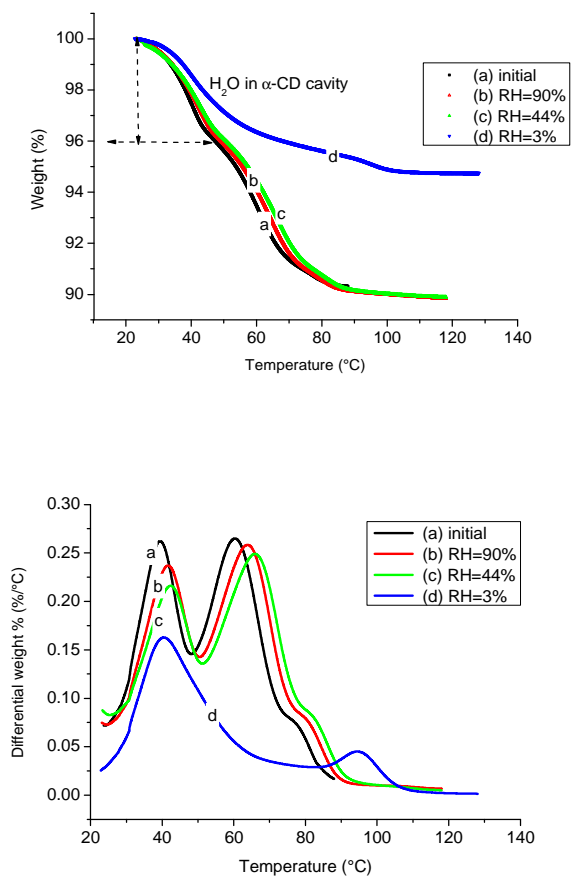

Figure 2 TGA scans on α-CD crystals previously equilibrated under different $RH$ values measured with a heating ramp of +1°C/min: a) Mass loss in wt. % versus temperature. The dash line indicates the fraction of water inside the α-CD cavity; b) differential of the mass loss.

Water retention in the first cases represents in molecular terms a total ~ 6 $H_2O$ molecules per α-CD. The maximum hydration of the cage structure is known as the hexahydrate structure (6$H_2O$ α-CD)[1], which is in good agreement with our determination from TGA measurements, showing that water saturation is stable in the range of *RH* values 44 to 90%. It is interesting to examine the dehydration steps by TGA: as shown by differential TGA in Figure 2b, the water loss appears as a stepwise process (see also Nakai et al. [23]). One can see three different steps with *RH*=44, 90% and initial state: the first dehydration step occurs between 27 and 48-50 °C, with a maximum of the derivative of the mass loss at 40°C. The water loss in the first stage is equivalent to 2.2 $H_2O$/α-CD molecules. According to Linert et al [26] the inner cavity of α-CD molecules contains up to 2 $H_2O$ molecules. Water molecules bonded in the cavity are less energetically stabilized and less fixed than those bonded outside, due to the particular molecular environment inside the cavity. The TGA experiments show distinctly that the thermal stability of the inner molecules is low. The stability of the outer 4 $H_2O$ molecules is divided in two categories: the less stable one desorbs between 50°C and 73-79°C, which represents 2.6-2.7 $H_2O$ molecules per α-CD and the most strongly bound water, 0.5 $H_2O$ molecule per α-CD, is lost between 78 and 93°C. The Monte Carlo simulations by Linert et al [26] do not allow precisely determining which water molecules outside the cavity correspond to the two distinct thermal stabilities that TGA experiments have clearly separated.

In the dry atmosphere, (*RH*=3%) the water content decreases to only 3 $H_2O$ molecules per α-CD and dehydration takes place in two steps: the major dehydration peak between 27 and 73°C starts at low temperature, similar to the first peak in the first samples, with a maximum at about 40°C. The peak broadens continuously above 40°C and no distinct second population appears, as before. A second dehydration peak is very stable: a small population of water molecules escapes between 85 and 110°C which was not seen in the previous samples (known to be the "cage" structure). The sample stabilized at low *RH* contains therefore a strongly bound water fraction, besides water inside the α-CD cavity, which is the less stable one. It is likely that the dry sample is in the "columnar structure", as established by Nakai et al. [23]. Besides, Hunt et al. [25] showed that the samples undergo a columnar-to-cage phase transition during vapor sorption. Rusa et al [20] noticed from DSC experiments that the α-CD columnar structure contains less water and supposed that the water molecules escaping first are primarily included in the "hydrophobic" cavity of α-CD. Their DSC thermograms were recorded at a high heating rate (+10°C/min) and the characteristic temperatures are shifted to much higher values than ours. Water molecules outside the cavity may be involved in hydrogen bonding with OH groups from both sides of the α-CD torus. In our experiments, we find a fraction of ~ 2.5 $H_2O$/α-CD molecules escaping in the first step within a broad temperature range. As this large peak (Figure 2b) involves more than 2 water molecules per α-CD, it possibly includes a loosely bound fraction of water outside the cavity which escapes in a cooperative way with the water inside the cavity in the supposed columnar structure. The second small peak in the high temperature range represents only 0.5 $H_2O$/α-CD molecules meaning that 1 $H_2O$ molecule bridges an average of 2 α-CD molecules and is tightly bound. This type of water did not appear in the previous states with higher *RH*s.

In conclusion, our experiments show that:

a) the thermal stability of the water molecules inside the cavity of the α-CD molecules (2 $H_2O$/α-CD) have a similar behavior, in the cage and the columnar structures, proving that the binding energy of the inner water molecules is mainly independent of the crystal structure. In the columnar structure, this fraction is the major part of the water content, in the cage structure, this is not the case.

b) The water molecules outside the cavity stabilize the structure of the crystals. In the cage structure, 4 water molecules are outside the cavity. By heating, the outside molecules escape in two steps, mainly between 50 and 73°C and between 73 and 95°C, the full dehydration is reached above 95°C (in our experimental protocol with a low heating rate). The outside molecules in the "cage" structure therefore have two different binding energies.

c) In the columnar structure, the water molecules inside and some molecules outside the cavity (total ~ 2.5 $H_2O/\alpha CD$) desorb cooperatively below 70°C. The full dehydration occurs at 110°. There is a very stable fraction of water that escapes between 85 and 110°C representing 0.5 $H_2O/\alpha$-CD.

### 3.2.2 β-CD

The same investigation was continued on β-CD microcrystalline powders. The results are shown in Figure 3a and b.

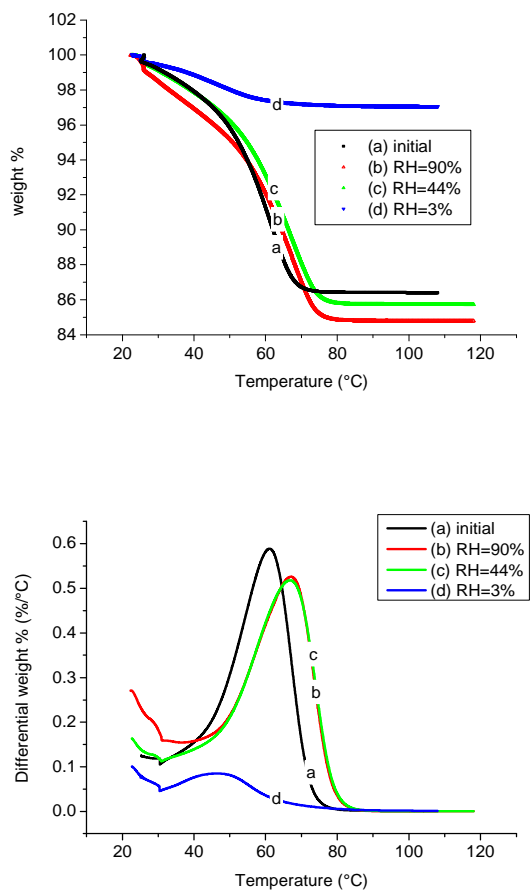

Figure 3 TGA scans on β-CD crystals previously equilibrated under different *RH* values measured with a heating ramp of +1°C/min: a) Mass loss in wt% versus temperature b) differential of the mass loss.

The water content of the crystalline powder varies strongly with *RH* (as shown in Table 1) the molecular ratio varying between 11.3 $H_2O$/β-CD and 1.9 $H_2O$/β-CD with decreasing *RH*. Sabadini et al [24] found in their sample 9.6 $H_2O$/β-CD and the largest values from the literature rise to 12 $H_2O$/β-CD [23].

One can see on the differential curves, Figure 3b, at the highest *RH*s, that all water molecules escape from hydrated crystals during heating between 40 and 80°C as a single broad population. The most hydrated samples (10.5-11.3 $H_2O/\beta$-CD) have the same differential curves, same thermal stability. No water molecules are left above 80°C, which is a much lower temperature than in α-CD crystals. It is agreed that 6-7 water molecules lie inside the β-CD cavity [26] and Saenger et al [8] proved that these water molecules and those in the crystal interstices are disordered and suggest that water molecules are mobile and that OH groups of β-CD may also rotate. By X-ray analyses on β-CD crystals exposed to different relative humidities, in the range of 100% to 15 %, they noticed that the water content reduces continuously from 12 to 9.4 molecules. In our case, the water content in the "as received crystals" is only 10 $H_2O/\beta$-CD and crystals are slightly less stable (Figure 3b) than in the fully hydrated state. We cannot distinguish different states of disorder of the water molecules in the crystal: in our TGA experiments there is a single desorption peak for all water molecules.

The β-CD sample stabilized at $RH$=3% has a very low water content (1.9 $H_2O/\beta$-CD) and loses all the water molecules before 70°C. Such a low hydration (2 $H_2O/\beta$-CD) was also observed by Nakai et al [23] for crystals equilibrated at $RH$=11% and the X ray diffraction showed a low crystallinity in this dehydrated state. Sabadini et al. [24] also noticed changes of X-ray diffraction patters between hydrated and dehydrated samples. However, the distribution of the water within the dehydrated β-CD crystals is not necessarily homogeneous [27], causing mechanical tensions which lead to cracks. According to Sabadini et al. [24] below $RH$=15% the crystalline structure collapses irreversibly. We did not perform X ray diffraction measurements on our samples, but the TGA measurements demonstrate that water escapes at low temperatures with no specific features. Al low $RH$, the remaining water molecules may not be involved into a crystalline structure: between the initial sate ("as received") and the dehydrated state at $RH$=3% the loss of water molecules is the largest one compared to the other CDs and can lead to a poor crystalline state.

### 3.2.3 γ-CD

The γ-CD powders were also analyzed by TGA after equilibration in different atmospheres. The results are shown in Figure 4.

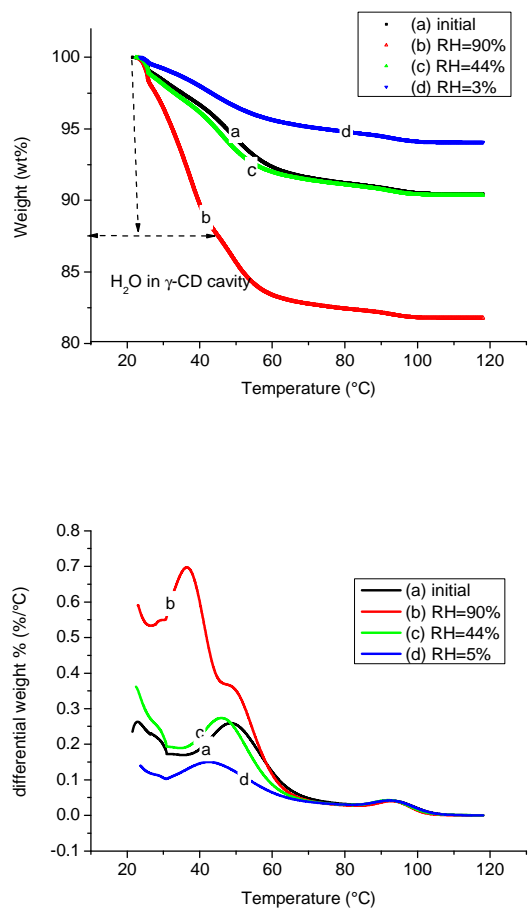

Figure 4 TGA scans on γ-CD crystals previously equilibrated under different *RH* values measured with a heating ramp of +1°C/min: a) Mass loss in wt. %. The dash line indicates the fraction of water inside the γ-CD cavity; b) differential of the mass loss.

γ-CD powders stabilized under various *RH*s exhibit different water retention levels (Figure 4a). Water content varies in these samples between 16 and 4.75 $H_2O$/γ-CD. Nakai et al [23] obtained in γ-CD recrystallized from water and stored at *RH*=93.6% crystals containing 17 $H_2O$ per γ-CD, which seems to be the highest possible hydration for these crystals. In our TGA experiments the powders start desorbing water as soon as they are out of the desiccator and the heating ramp starts. The beginning of the dehydration cannot be very accurately determined. The differential TGA curves are shown in Figure 4b. Our most hydrated sample 16 $H_2O$/γ-CD (*RH*=90%) exhibits a three step dehydration process: the first step is centered around 36°C, from room temperature to 46°C and corresponds to desorption of ~ 10 $H_2O$/γ-CD. According to [24] and to earlier publications, 8.8 $H_2O$ molecules are present in γ-CD cavities in hydrated forms. Besides, Ding et al [28] found by neutron diffraction studies that 15.7 $H_2O$/γ-CD crystals are arranged in the "cage" type structure. Therefore, our first dehydration peak in Figure 4c may correspond to desorption of water molecules inside the γ-CD cavity in the "cage" structure of the crystal. The thermal stability range of this peak is comparable to the α-CD cavity, which indicates the same energetic state. The second peak, overlapping the first one, could be the interstitial water of the "cage" structure with a fraction of 4.7 $H_2O$/ γ-CD and finally another very stable fraction 1.3 $H_2O$/γ-CD desorbs between 80 and 108 °C (*RH*=90%).

γ-CD samples stabilized at *RH* =44% and in the initial state ("as received" sample) contain less water (7.6 $H_2O$/γ-CD) and exhibit a two-step dehydration process: the main dehydration peak between 42 and 49°C involves the desorption of 6.8 $H_2O$/γ-CD and the second one between 80 and 108 °C containing ~0.8 $H_2O$/γ-CD which in the very stable fraction. These crystals may have a different structure: by sorption or desorption experiments of γ-CD powders it was observed [23] a stable form with a composition of 7 $H_2O$/γ-CD, which had a different diffraction pattern compared to either 17 $H_2O$/γ-CD and vacuum dried samples (4-5 $H_2O$/γ-CD). It is not clear which type of crystalline structures are generated by stabilizing the γ-CD powders in the desiccator at medium and low *RH*. Additional X ray diffraction studies should be performed to elucidate the different states. The main dehydration peak in our differential TGA curves (initial sample and *RH*=44%) should correspond to desorption of the water inside cavities. The γ-CD sample in the desiccator $RH$=3% contains 4.7 $H_2O$/γ-CD and the thermal stability is less than the previous cases. Interestingly, all samples had a similar (~1$H_2O$/γ-CD), very stable, fraction of water molecules desorbing at high temperatures.

In conclusion, TGA experiments, combined with the initial stabilization of the microcrystalline powders in various $RH$ atmospheres, provide a very simple, quantitative, analysis of the degree of hydration and of the phase type in CD crystals.

It was found first of all that the water content of the crystals is very sensitive to the storage conditions. Two basic structures of CD crystals have been previously identified by X ray diffraction, namely "cage" and "columnar" structures for each CD. The two structures can transform reversibly from one into other form during water vapor adsorption/desorption.

TGA experiments locate the water molecules in hydrated crystals mainly inside the cavity with a content which depends on the size of the cavity, the degree of hydration and the crystalline structure. The characteristics of hydrated crystals and the fraction of water evaporating during the first peak in our TGA experiments are summarized in Table 2.

Table 2

Characteristic hydration of CD crystals and determination of the first dehydration peak by TGA for samples equilibrated at *RH*=90%.

| CD | Maximum hydration of crystals | $H_2O$ inside cavity | Crystal Hydration (*RH*=90%) | 1st peak TGA (*RH*=90%) |
|---|---|---|---|---|
| α-CD | 6 $H_2O$/α-CD[1] | 2 $H_2O$/α-CD | 6 $H_2O$/α-CD | ~2.2 $H_2O$/αCD |
| β-CD | 12 $H_2O$/β-CD[23] | 6 $H_2O$/β-CD | 11.3 $H_2O$/β-CD | ~11 $H_2O$/β-CD |
| γ-CD | 17 $H_2O$/γ-CD[23] | 8.8 $H_2O$/β-CD | 16 $H_2O$/γ-CD | ~10 $H_2O$/γ-CD |

The molecules which desorb during the heating ramp in the first peak are mainly those inside the cavity, but it seems that a fraction of the molecules from outside also desorb cooperatively allowing a rearrangement of the whole crystalline structure. In the most hydrated α-CD and γ-CD crystals water escapes in three steps. In β-CD crystals all water molecules desorb within a single step, below 80°C showing the most weakly bound structural water. It was found a small, very stable fraction, 0.5 -1 $H_2O$/CD, strongly bonded the CDs columnar structure.

The interpretation of desorption temperatures provides a clear description of the role of water in stabilizing the hydrated crystals. The TGA experiments help to discriminate between the weakly and strongly bound sites of the water molecules, which may be analyzed further by more refined techniques. The TGA experiments with stabilization in controlled atmospheres may be developed also for inclusion complexes of CD crystals. Water desorption should be related to the guest molecule, the composition of the crystals and their structure.

Another aspect probing water mobility inside hydrated crystals is their solubilization in water relation with the molecular composition of the crystals, the concentration in solution and the temperature. The next section deals with the solubilization of the CD crystals using µDSC and DSC techniques.

## 3.3  µDSC and DSC

The aim of this investigation is characterize the dissolution and melting of the CD crystals. The experimental work was performed with "as received" crystals. All the microcrystalline powders have a residual hydration, but, from the previous results, they are not in the same crystalline form: α-CD and β-CD were initially in the "cage" structure, while the γ-CD was mainly in the "columnar" structure. It was established [22] that the main difference in enthalpy of dissolution appears between hydrated and non-hydrated crystals, but the influence of the crystal structure was not reported .

All experiments were performed with a slow heating rate (unless otherwise stated) +0.1°C/min.

### 3.3.1  α-CD

We measured the heat flow during heating between room temperature and 90°C. In Figure 5 the concentration of the powder varies from 10 to 30 wt. %.

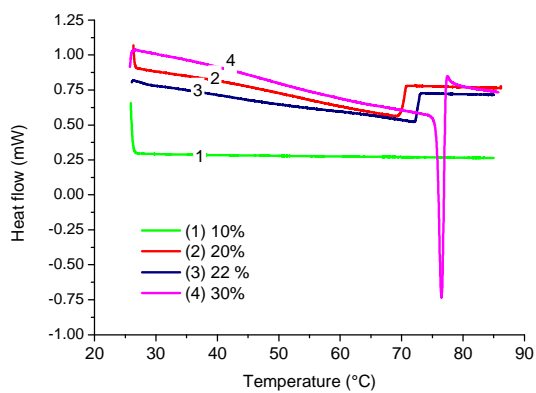

Figure 5 Heat flow of α-CD dispersions with increasing concentrations: $c$=10, 20, 22 and 30wt%. The melting peak appears for the most concentrated sample.

As expected, the trace of the heat flow at the lowest concentration 10 wt. % α-CD shows no enthalpy changes, because the powder is already fully dissolved at room temperature. The samples with higher concentrations 20 and 22 wt.% exhibit distinct broad endothermic signals with an abrupt end and return to the base line at a temperature which depends on concentration: $T$=69°C at $c$=20 wt.% and $T$=72°C at $c$=22 wt.%. The heat input is necessary to dissolve the powder between approximately $c$=14 wt. % at $T$=25°C and $c$=22 wt. % at $T$=72°C, in agreement with the previous publications. The total surface of the broad peak represents the enthalpy of dissolution for each concentration. When concentration reaches $c$=30 wt.%, one can follow Figure 5, in first place the heat flow for dissolution, below 75°C, followed by a very sharp, narrow peak which reveals the melting transition when the solution becomes supersaturated; the peak is centered on $T$=76.5 ±0.5°C. This peak remarkably shows that the solid α-CD can melt in the supersaturated solution as a distinct phase. At this particular temperature, the melt (fluid), and the solid α-CD coexist with the saturated aqueous solution containing dissolved α-CD molecules. Water molecules also exchange between the hydrated solid, the liquid α-CD and the saturated solution. The presence of an excess of water in a liquid state allows equilibration of the chemical potentials of the water molecules between the hydrated solid, the hydrated liquid, the aqueous solution and finally the vapor in the cell. This equilibrium is achieved at a precise temperature: if this temperature characterizes the melting temperature of α-CD hydrated crystals, then the temperature is independent on the solid concentration and the enthalpy per unit mass undergoing the melting transition is a constant. Therefore, we continued to increase the powder concentration in solution and we observed the melting peaks as shown in Figure 6 until 100 wt. % (no addition of water).

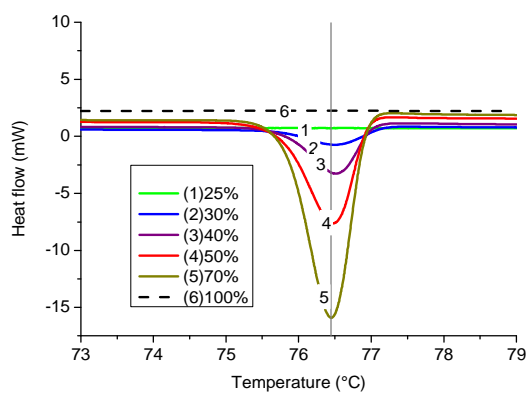

Figure 6 Heat flow for the very concentrated α-CD solutions showing the melting peaks in the supersaturated ones. The melting temperature is identical, the area of the peaks increases with α-CD concentration

The sharp peak temperature is always centered at $T_{melt}$= 76.5 ±0.5°C which is the melting temperature of α-CD hydrated crystal. Above 77°C the α-CD is in the liquid state. At 100 wt. %, which is the initial, "as received" crystal, without any additional water, there is no melting transition. At concentration of $c$=80wt%, the melting peak is visible, the heat flow saturated at the peak temperature (-20mW), at higher concentrations ($c$=85-90 wt. %) the powders were not homogeneously dispersed (data not shown). The analysis of the solubilization and the melting enthalpies is presented in Section 4.

### 3.3.2 β-CD

The β-CD crystals are not very soluble at room temperature, but dissolution improves upon heating.

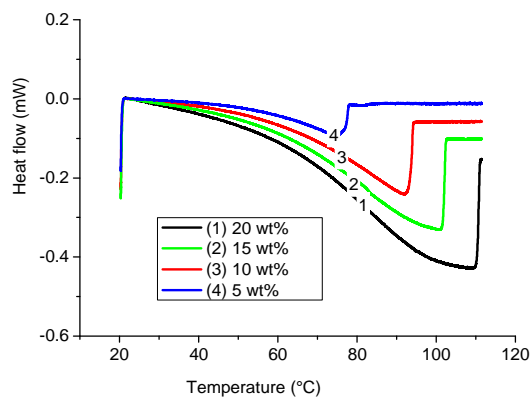

Figure 7 Heat flows for β-CD solutions with increasing concentrations. The base lines were shifted to zero initial value to facilitate their comparison.

Heat flows are shown in Figure 7 for solutions with increasing concentrations between $c=5$ and $c=20$ wt. %. The initial value of the heat flow was set to 0 in order to more easily compare the traces for the different concentrations. Dissolution is an endothermic process, as for α-CD and enthalpy increases with temperature: $c=5$ wt. % totally dissolves at $T=76°C$, $c=10$ wt. % at $T=93°C$, $c=15\%$ at $T=102°C$ and $c=20$ wt. % at $T=110°C$. The heat flow increases strongly with the temperature during the heating ramp with a strong curvature: the area between the base line and the heat flow trace determines the dissolution enthalpy versus temperature. The maximum temperature corresponds to dissolution of crystals between the initially soluble concentration and the total concentration of the dispersion. No saturation appears in the rage of temperatures accessible with this equipment (maximum 115°C) and the melting point of the β-CD crystals is not reached.

Additional experiments were performed with DSC Q2000 from TA Instruments, the heating rate was higher, +0.5°C/min. Steel cells, with 110µl volume, with hermetic lid allowing a maximum pressure of 25 bars, were used and prevented water evaporation during heating ramps from 25 to 200°C. The heat flow curves (Supplementary Information) with concentrations between $c$=40 wt. % and 80 wt. % are similar to those presented in Figure 7 and show that β-CD continues to dissolve in water. The most concentrated sample totally dissolved at ~155°C and the heat flow returned to the base line. We did not try DSC experiments with larger concentrations (>80 wt. %). We did not observe the melting transition of the solid in the whole concentration range.

### 3.3.3  γ-CD

The γ-CD crystalline powders are dispersed at room temperature in the µDSC cell and submitted to heating at constant rate, after equilibration. The heat flow traces are shown in Figure 8, vertically shifted at the beginning of the ramp in order to start at a value close to zero and compare the data. The range of concentrations is $c$=50 to $c$=80 wt. %. The heat flow is again endothermic. The dissolution shows a gradual saturation: for instance, at c= 50 wt.% and $c$=60 wt.% dissolution ends progressively between 55°C and 72°C, while a clear end step was noticed in α-CD and β-CD crystals. The melting temperature of the crystals in the saturated solution is also observed in the most concentration solution, c=80 wt. % with a sharp peak centered at $M_{elt}$=74.8°C.

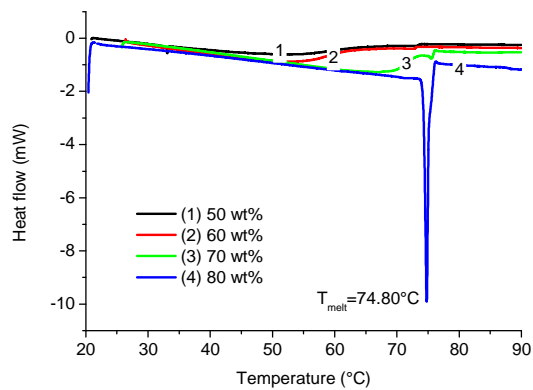

Figure 8 Heat flows for γ-CD solutions with increasing concentrations, between 50 and 80 wt. %. The base lines were shifted to zero initial value to facilitate the comparison.

It is interesting to better examine the melting peak in a narrow temperature range. This graph is shown in Figure 9. Obviously, here two melting peaks coexist: the main peak is at $T_{melt}^1 =$ 74.8 C and a second small peak is centered on $T_{melt}^2$=75.6°C. The small peak is visible on the trace at $c$=70 wt. %, while the largest one appears only at highest concentration $c$=80 wt. %. We suggest a possible interpretation of the two populations: the initial composition of γ-CD (7 $H_2O$/γ-CD) is below the maximum hydration (17 $H_2O$/γ-CD) as discussed in Section 3.2.3. Therefore two structures with different hydration and may exhibit different melting temperatures, with a small difference of the order of 1°C. These temperatures are also very close to the α-CD melting temperature in "cage" structure. The dissolution traces may also reveal a difference of the hydration of the crystalline structure: the large temperature spread of the end of the dissolution range (~ 10°C) could possibly also be related to the different phases in the powder. In order to elucidate these points one should first stabilize the powders in very humid atmospheres ($RH$=90%) in order to generate the most hydrated structure (17 $H_2O$/γ-CD), before starting the heating ramp with µDSC experiments.

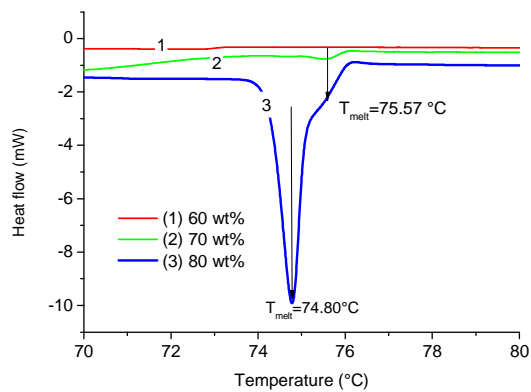

Figure 9 Heat flows of γ-CD concentrated solutions showing two melting peaks around 75°C. The difference between the two melting temperatures is less than 1 °C.

The dissolution and melting of γ-CD crystals exhibit similar behavior to α-CD. The range of concentrations is however much broader, due to a larger solubility of the powders at room temperature. The presence of two different molecular arrangements is suspected from the μDSC measurements.

# 4 Discussion

The μDSC and DSC determines of dissolution and melting of the CDs in aqueous solutions. We shall examine more deeply the results.

## 4.1 α-CD

The dissolution enthalpy of α-CD can be derived from the first part of the ramp, below $T=75$°C corresponding to dissolution of a concentration of 25-13.5=11.5 wt. %. To evaluate precisely the enthalpy of dissolution, it is important to control the initial temperature and dissolution time within the μDSC cell, before starting the heating ramp. This caution ensures that the solution is in equilibrium. We performed a rigorous experiment with a total concentration of $c=25$ wt. %. The total enthalpy of dissolution was measured after integration of the peak between 25 and 75°C and then normalized with the mass of α-CD solubilized. We derived from this experiment the dissolution enthalpy:

$$\Delta H_{dissolution}(\alpha\text{-CD}) = 38.5 \pm 1 \text{ J/g or } 37.4 \pm 1.1 \text{ kJ/mole}$$

This value has the same order of magnitude compared to the literature [21], 32 J/g, but is smaller. Besides, the integration of the enthalpy peak at any temperature, between $T=25°C$ and $T=75°C$ normalized by the specific dissolution enthalpy (in J/g), determines the solubility curve for α-CD, expressed by wt. % versus temperature. The solubility curve is reported Figure 10.

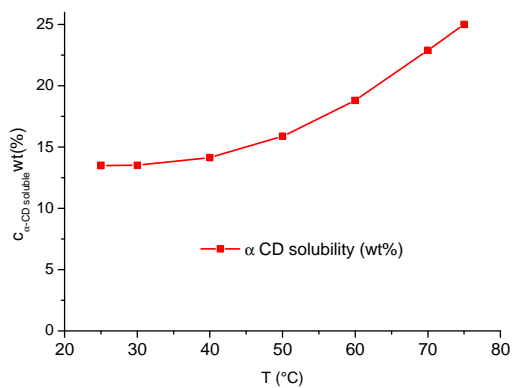

Figure 10 Solubility of α-CD versus temperature derived from µDSC measurements on a solution equilibrated at 25 °C, prior to heating at a constant rate +0.1°C/min. The solution concentration is $c$=25wt%.

The solubility from Table 1 in [21] differs from ours: their data and ours both establish the same maximum dissolution concentrations (*circa* 25 wt.%), but in a different temperature range. Jozwiakowski and Connors [21] observed a steady increase of the dissolved mass with temperature until $T=48°C$, but they do not report any measurements at higher temperatures. In our case, the data shows that dissolution of the powders proceeds until when the melting of the crystals starts at a fixed temperature, as discussed below.

The area of the melting peaks increases as expected with α-CD concentration above c=25wt% due to the non-soluble fraction of crystals in supersaturated solutions. The area under the sharp peak is the melting enthalpy of the crystals. In Figure 11 the area of the peaks expressed in J is plotted versus total concentration of α-CD in the dispersion. The area increases linearly with the concentration and the saturation concentration is determined by extrapolation at zero enthalpy.

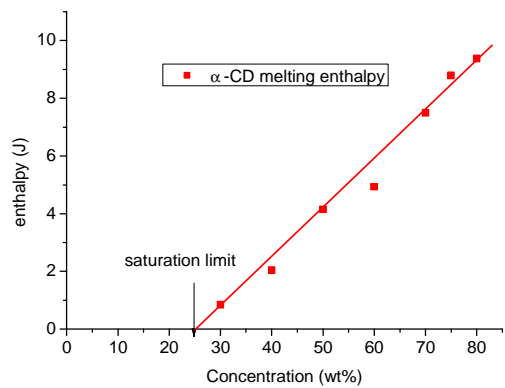

Figure 11 Melting enthalpy plotted versus the solution concentration for α-CD crystals. The saturation concentration is the value extrapolated to zero enthalpy.

One can deduce the melting enthalpy by normalizing the enthalpy by the mass of non-soluble α-CD:

$$\Delta H_{\text{melt}}(\alpha\text{-CD}) = 21 \pm 1 \text{J/g}$$

and the saturation limit is:

$$c_{\text{satur}}(\alpha\text{-CD}) = 25 \pm 1 \text{wt\%}$$

The melting enthalpy of the hydrated α-CD crystals is smaller than the dissolution enthalpy. The experiment shows for the first time to our knowledge that α-CD can be obtained in the liquid phase above T=78°C in equilibrium with the aqueous solution.

## 4.2 β-CD

The dissolution enthalpy of β-CD crystals can be calculated from the heat flow traces presented Figure 7 for different dispersions. There is a linear increase of the area under the dissolution ramp with the mass of crystals in the dispersion. The enthalpy of dissolution is:

$$\Delta H_{dissolution}(\beta\text{-CD}) = 41 \pm 1 \text{ J/g or } 46.5 \pm 1.1 \text{ kJ/mole}$$

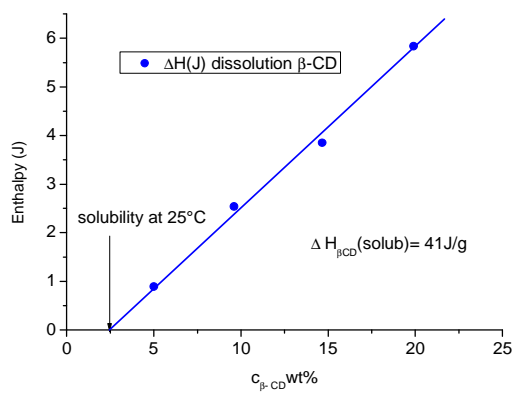

Figure 12 Enthalpy of dissolution β-CD crystals plotted versus the concentration. The extrapolation to zero enthalpy indicates the initial solubility of the β-CD crystals at 25°C.

From dissolution enthalpy, one can derive the solubility curve versus temperature, as shown in Figure 13. The data from [21] is shown for comparison. Our results cover a larger domain of concentrations. The enthalpy of dissolution given in [21] is 30.60 J/g (from Table V) which is much smaller than our result.

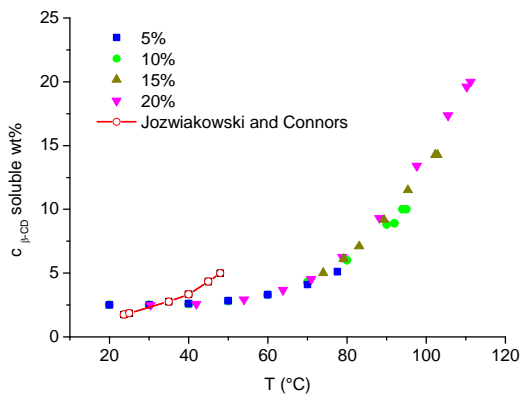

Figure 13 Solubility of β-CD crystals versus temperature, derived from heat flow measurements at several concentrations indicated on the graph. The results from [21] are shown for comparison.

We did not reach the limit of solubility of β-CD crystals: $c$=80 wt. % was totally solubilized at 160°C. If β-CD could show a melting temperature, it would be above 160°C. Larger concentrations require a very precise weighing of the sample and control of water evaporation is critical. The melting temperatures of α and γ-CD are close to each other, in the range 75 to 76°C. The presence of water in excess is a key factor in controlling the thermal properties of the solid CD. These aspects have not been noticed in the previous publications with calorimetry methods or others.

The dissolution enthalpy of β-CD is larger than for α-CD, as already established in previous work, although our numerical values differ from the existing literature.

Coming back to the context of complex formation, in drug delivery applications, thermo-analytical [29] methods are generally used to detect whether the guest substance undergoes changes (like melting, evaporation[30], decomposition,[31] oxidation or polymorphic transition) before the thermic degradation of CD occurs. Such changes of the guest substance can be related to the complex formation with CD. The nature of the drug, of the CDs used and of the method of preparation of complexes have been found to influence considerably such changes[29].

Our approach suggests a different method of characterizing of the CD complexes through the use of thermo-analytical methods: starting from CD crystal and looking for the changes of solubility and thermal transitions, due to complex formation, one can follow accurately the properties of the crystals after inclusion formation, provided the solvent is present in solution or when the solids are equilibrated in controlled humid atmospheres. In view of these characterizations, the concentration of the solid phase should be increased by centrifugation of the suspensions, which is very efficient in separating the solids, due to the large difference of density between the CD crystals and liquid phase. DSC and TGA do not require large quantities of sample (10-50 mg); a larger quantity of solid (100mg-1g) is necessary for micro DSC. Quantitative determination of changes of the thermal properties of the hydrate crystals when CD complexes are built should help to characterize the complexes at least for model systems.

## 5  Conclusion

This work shows various thermodynamic aspects of dehydration, dissolution and melting of CD crystals. New quantitative results illustrate the large differences between these molecules belonging to the same family of cyclic oligomers. Hydration is intimately involved in all the molecular arrangements. Water is a distinct, labile component in the crystalline solids, which can be removed or exchanged during sorption, desorption or melting transitions. It is shown in our study, for the first time that the melting of the CD crystals in a separate liquid phase can be observed provided that the hydrated solid is in contact with the saturated aqueous solution. This study is a preliminary step to understand the stability of the complexes that CD can form with a large number of guest molecules. Water is one of the (small) guest molecules in CD cavities. It is supposed in general (see for instance ref. [7]) that complex formation is "evidently reversible", but the conditions to observe the reversibility of the complexes in controlled environments are not well understood. Our investigation may open a new field for analyzing either complex formation, solubilization or desorption in controlled environments.

## Acknowledgments

We are grateful to TA Instruments, in particular to Jacques Loubens, for performing the DSC experiments on β-CD samples and for his help and precious advice. M.D. and K.B. thank Dr. Sylviane Lesieur the fruitful the discussion and for her expert advices.

## Supporting information available

Dissolution of α-CD crystals by optical rotation measurements; SEM images of α, β and γ-CD crystals; heat flow obtained by DSC for β-CD crystals in solution with concentrations between 40 to 80 wt.% and temperature range between 25 and 200°C. This material is available free of charge via the Internet at http://pubs.acs.org.

## Author information

Corresponding author:

Madeleine Djabourov, E-mail: Madeleine.Djabourov@espci.fr  Tel: 33 1 40 79 45 58

## Author Contributions

The manuscript was written through contributions of all authors. All authors have given approval to the final version of the manuscript.

**Note**

The authors declare no competing financial interest.

# Table of contents graphics

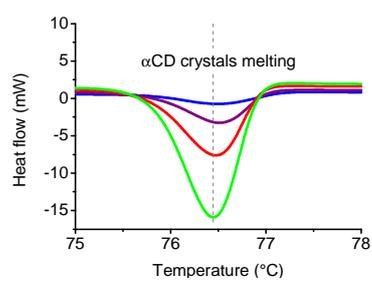